# SArF Map: Visualizing Software Architecture from Feature and Layer Viewpoints

Kenichi Kobayashi, Manabu Kamimura, Keisuke Yano, Koki Kato, Akihiko Matsuo
Software Systems Laboratories
Fujitsu Laboratories Ltd.
Kawasaki, Japan
{kenichi, kamimura.manabu, yano, kato_koki, a_matsuo}@jp.fujitsu.com

*Abstract*—To facilitate understanding the architecture of a software system, we developed *SArF Map* technique that visualizes software architecture from feature and layer viewpoints using a city metaphor. SArF Map visualizes implicit software features using our previous study, SArF dependency-based software clustering algorithm. Since features are high-level abstraction units of software, a generated map can be directly used for high-level decision making such as reuse and also for communications between developers and non-developer stakeholders. In SArF Map, each feature is visualized as a city block, and classes in the feature are laid out as buildings reflecting their software layer. Relevance between features is represented as streets. Dependency links are visualized lucidly. Through open source and industrial case studies, we show that the architecture of the target systems can be easily overviewed and that the quality of their packaging designs can be quickly assessed.

*Index Terms*—Software visualization, city metaphor, software architecture, program comprehension, software clustering, dependency graph.

## I. Introduction

Understanding software architecture is one of the important steps in software development and maintenance activities. To recover lost or outdated knowledge, software visualization techniques and tools have been often utilized in various software understanding contexts [1][2][3].

There are many existing studies that visualize the explicit source code organizations such as packages and directories [4][5][6][7]. To deeply understand the software architecture, we suggest that it is necessary to reveal the implicit information of software architecture, *features*. In this paper, we use the term *feature* as the definition in the feature location studies: A (software) feature is a functionality of the system that can be triggered by an external user [8]. If such implicit information is visualized, it is expected to greatly facilitate understanding software systems. Therefore, we hypothesized that features can be collected using an appropriate software clustering algorithm.

We propose a novel software visualization technique, *SArF Map* on the basis of the hypothesis. "SArF" is the abbreviation of "Software Architecture Finder." It visualizes software architecture from a feature and layer viewpoints using a city metaphor [9]. Features are extracted by our previous study, SArF software clustering algorithm [10]. It successfully gathers classes that share a common feature into the same cluster (in this paper, a *class* is compatible with a source file and any unit of software entity).

Since features are high-level abstraction units in software systems, it is expected that users of SArF Map can make high-level decisions on the target system. Besides, it is interpretable for non-developer stakeholders and can be used as a common mental model of the target system among various statuses of stakeholders. Since SArF Map is based on the software dependency graph, the knowledge and insights extracted from the map can be directly available for various software development and maintenance activities such as change impact analysis and reuse.

To visualize the two different viewpoints, our visualization consists of two different layouts. In the city metaphor we used, a feature is represented as a city block, and layers are represented as slopes and north-south direction. Laying out features is laying out streets, and laying out layers is laying out buildings in a city block. We called our layout scheme, the *Street and Block Tree Layout*. Relevant features are connected by streets and located in close positions in order to facilitate the understanding of the architecture from a feature viewpoint. By comparing packages with features and layers in a map, the package organization can be quickly assessed through color patterns appeared there.

SArF Map is fully automated by virtue of the high degree of automation of the underlying SArF algorithm [10]. The output is deterministic, and the input is only jar files or a dependency graph. It is language-independent. Because of these characteristics, SArF Map is highly applicable.

The remainder of this paper is organized as follows: we will show related work in section II. Our novel software visualization technique, SArF Map will be explained in section III. In section IV, we will set up research questions. In section V, we will show case studies and assess the research questions. The threats to validity of SArF Map will be discussed in section VI. Finally, we will summarize in section VII.

## II. Related Work

Software visualization has been studied for various goals [1][2][3], such as to understand the overview of large software systems and their architectures [4][5][6][7] and to understand semantic concepts [11].

A city metaphor is widely adopted in many studies [5][6][7][9][12][13], it is intuitive and navigable, and it can represent various software structures and metrics at the same time [3]. We employed it for the same reason. Caserta and Zendra pointed out that one drawback of a city metaphor is it can be laid out on only a 2D space [3], but we take advantage of the restriction. We perform our blank visualization in a pure 2D space as a blank map and use it as a shared mental model between various statuses of stakeholders.

CodeCity [5][14] is one of approaches using a city metaphor. It maps a class to a building and maps a package to a city district using a hierarchical rectangle packing tree map layout. It can display various metrics simultaneously and represent a source code hierarchy naturally, and it is highly scalable. The most important difference between SArF Map and CodeCity (and other existing city metaphor visualizations) is that a city block in SArF Map is a feature, which is implicit information that can be recovered only using software clustering.

Software clustering is a technique which decomposes a given system into several subsystems or groups of classes with manageable sizes. Bunch [15] is a graph clustering approach finding high-cohesion and low-coupling modular clusters. SArF [10] is also a graph clustering approach that gathers classes that share a common feature into the same cluster. Semantic information such as identifiers and comments in source code are used for natural language processing clustering techniques [11]. To understand behavioral properties of a system, dynamic information such as execution traces is used for clustering [9].

Scanniello et al. [16] extracted software layers. Their technique needs some human decisions and is not applicable to software with complex layer structures such as Fig. 11(a). Since SArF Map decomposes software to features, complex layer structures are untangled, and thus layers are automatically visualized. It is further described in subsection III.D.1).

The Street and Block Tree Layout (SBTL) is an extension of the hierarchical street layout used in previous studies [7][13]. SBTL layouts city blocks instead of buildings and employs an additional layout algorithm to layout the buildings in a city block. Although one of the weak points of the hierarchical street layout is low area efficiency, SBTL has relatively high area efficiency, because many buildings are packed into a city block.

Visualization tools have been used with software clustering since early times. For example, Bunch used graphviz tool (http://www.graphviz.org/) to assess the clustering results. However, tightly coupled combinations of software clustering and visualization for large-scale software have rarely appeared. To our knowledge, only one semantic-based software clustering approach, Software Cartography [11], exists.

## III. SArF Map Visualization

In this section, we first present an overview of SArF Map technique. Next, we present its procedure. Then, we briefly explain its underlying software clustering algorithm, and then, we describe the details of the steps in the procedure. Finally, we describe the supporting system.

### A. Overview

Our proposed technique, SArF Map, visualizes the overall view of target software from feature and layer viewpoints as a map. We also call a generated map a SArF Map. Figure 1 is an example SArF Map of Weka 3.0, an open source data mining tool [17]. Using this example, we will explain the concepts and algorithms of SArF Map in the following subsections.

To visualize software architecture, we employed a city metaphor. Table I shows the mapping between entities and concepts in concrete software and metaphors in visualization used in SArF Map.

The SArF Map in Fig. 1 shows 142 colored buildings and eight city blocks. Each building is a class, and each block is a group of classes implement some feature. The words over the buildings are the keywords in the feature. For example, the upper left city block with green buildings is characterized as a word "estimator", and its feature may be an estimator, one of common functionalities of data mining tools.

The streets in the map represent the relevance between features, and the colored curves represent the dependencies between classes. The detailed explanation is given in the later subsections.

TABLE I.  ENTITY AND CONCEPT MAPPING IN CITY METAPHOR

| Entity and Concept | Metaphor in Visualization |
| --- | --- |
| class / source file | building |
| feature (cluster) | city block |
| layer | north-south way and slope |
| relevance between features | street and distance |

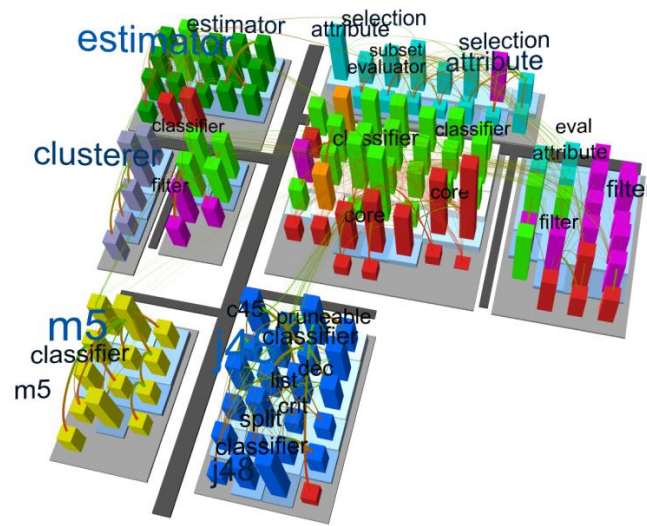

Figure 1.  SArF Map of Weka 3.0 (142 classes, 9 packages)

The input data for SArF Map is only a dependency graph. In this paper, it is extracted from the jar files of the target software. The extracted dependencies are method calls, field accesses, inheritances, and class type references. After clustering is performed, all member-level dependencies are summed up to class-level dependencies. Therefore, generated maps and their users only treat class-level dependencies.

SArF Map technique is fully automated. All examples and case studies in this paper are generated automatically.

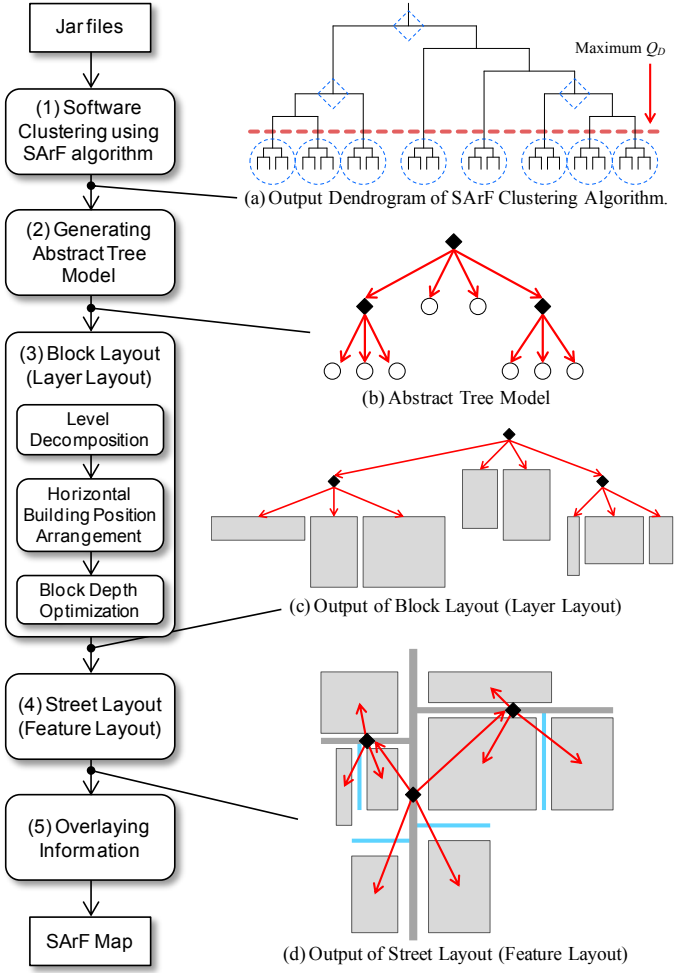

Figure 2. Procedure of SArF Map and Intermediate Data

## B. Procedure

The procedure of SArF Map technique comprises the following steps as shown in Fig. 2:

(1) Perform software clustering on the target software system using SArF clustering algorithm, which gathers features in the resulting clusters. The output is a dendrogram as shown in Fig. 2(a). Subtrees enclosed blue dashed circles are clusters. The detail is described in subsection III.C.
(2) Generate the abstract tree model of features as shown in Fig. 2(b) from the dendrogram. Clusters are converted to leaf nodes as features. Branch nodes in the dendrogram are expanded in such a way that closer nodes are divided later.
(3) Lay out the buildings (classes) in the city block for each feature cluster as shown in Fig. 2(c). The layout algorithm reflects the layer of each class on its position. The detail is in subsection III.D.

(4) Lay out features on the streets. Tree branches are represented as streets. Blocks of relevant features are closely placed as possible. In this step, a blank map as a 2D mental model for users is realized as shown in Fig. 2(d). The detail is in subsection III.E.
(5) Overlay information used for the analysis on the blank map. The detail is in subsection III.F.

## C. SArF Software Clustering Algorithm

The underlying software clustering algorithm of SArF Map is our previous study, SArF algorithm [10], which gathers classes implementing the same feature or feature set into an identical cluster. In this subsection, we briefly explain the algorithm.

A software feature is implemented as a set of classes, which interact with each other to realize the feature. Such interactions can be represented as some sort of dependency graph. We assumed that the likelihood that two classes share a common feature can be quantified, and we also assumed that when the likelihood is assigned to the weight of the interaction, subgraphs with dense weighted edges in the dependency graph tend to contain classes sharing the same feature. Therefore, the problem finding features can be considered as weighting edges and finding dense subgraphs. SArF algorithm consists of two components, the Dedication Score [10] to weight edges and a Modularity Maximization [18][19] to find subgraphs.

Modularity Maximization is one of graph clustering or community detection algorithm used in social network analysis, biochemical network and so on [18]. We used its weighted and directed extension [20]. To find clusters, it maximizes the objective function, Modularity $Q_D$ [20], which means the sum of the significance of the edges in clusters. The significance is defined as the difference between the actual weight of an edge and the expectation of its weight.

The Dedication Score is the importance of the dependency relationship based on fan-in analysis. Figure 3 shows simple examples. If only class A depends on class X, X is probably dedicated to A to implement the same feature. On the other hand, Y can hardly be said that Y is dedicated to $B_1$. Thus, we defined the Dedication Score of the dependency ($x \rightarrow y$) as 1/fan-in($y$). In practice, the Dedication Score considers the class-method hierarchy, and its definition is more complex [10]. However, this simple explanation is enough to understand its basic concept and its interpretation.

SArF algorithm performs better than former studies and successfully extracts features [10]. The output of SArF algorithm is clusters (groups of classes) and a dendrogram.

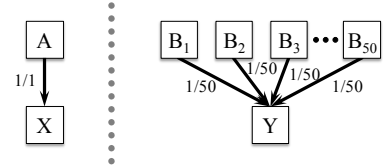

Figure 3. Dedication Score (Simple Case)

## D. Block Layout Algorithm for Layers

Software is often totally or partially layered. Well organized software tends to have good layers in its architecture. Since the dependencies between layers are one-way, recognizing layer structure makes software comprehension tasks simpler. By recognizing the absence of layer structures and the existence of cyclic dependencies, the maintainability of the software can be assessed.

We visualized layer structures by laying out classes in a city block. To make it easier to identify layers, all classes are grid-aligned. As shown in an example in Fig. 4, the layout procedure consists of the three steps, (1) *Level Decomposition*, (2) *Horizontal Building Position Arrangement*, and (3) *Block Depth Optimization*, described in the following subsections.

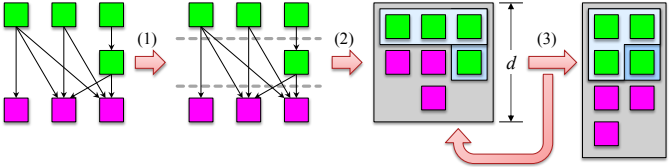

Figure 4.  Example of Block Layout Procedure

*1) Level Decomposition:* At the first step in Fig. 4, the dependency graph of the classes in the cluster is decomposed to some levels. In this paper, a level is defined as a fine-grained decomposition of a layer as shown in Fig. 5 and is a group of classes in almost the same distance from the top entries or the bottom exits in the block. Determining a layer is not trivial, but determining a level can be automatically performed as follows.

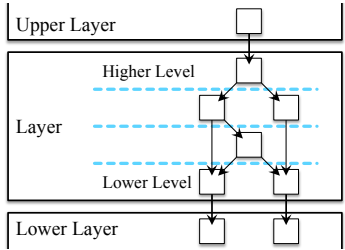

Figure 5.  Layers and Levels

The desired properties of a level is that it depends on only lower levels and that it is depended on by only upper levels. Unfortunately, practical software often has cyclic dependencies, thus ordinary topological sort algorithms cannot be applied to identify levels. We developed the Greedy Level Decomposition algorithm shown in Fig. 6. In the line 11, we introduced the classic Greedy Cycle Removal heuristic [21] to cope with cycles. By applying the algorithm, the dependency graph is successfully decomposed to some levels as shown in Fig. 4.

*2) Horizontal Building Position Arrangement:* At the second step in Fig. 4, each level is packed into the rectangle block in the descending order. Each building (class) is placed at the position horizontally close to its predecessor neighbors and successor neighbors as possible. In order to do that, X-coordinate of the building is calculated to minimize the following energy function:

$$f(x_i \mid i) = \sum_{j \in \text{ins}(i), l_j \ne l_i} \left( d_{ji}(x_i - x_j) \right)^2 + \sum_{j \in \text{outs}(i), l_j \ne l_i} \left( d_{ij}(x_i - x_j) \right)^2, \quad (1)$$

where $i$ and $j$ are buildings, and $x_i$, $l_i$, $x_j$, and $l_j$ as X-coordinates

```
1:# G is the set of the input nodes.
2:func GreedyLevelDecomposition(G) {
3:    UpLevels, LowLevels := ()  # List of Sets
4:    while (|G| > 0) {
5:        # ins(n) returns the set of predecessor nodes of n in G
6:        UL := {n in G where |ins(n)| = 0}  # Set
7:        # outs(n) returns the set of successor nodes of n in G
8:        LL := {n in G - UL where |outs(n)| = 0}  # Set
9:        if (|UL| + |LL| = 0) {
10:           # Greedy Cycle Removal heuristic
11:           UL += argmax(n in G) (|outs(n)| - |ins(n)|)
12:       }
13:       for c in UL + LL {
14:           G -= c
15:           for n in outs(c) { ins(n) -= c }
16:           for n in ins(c) { outs(n) -= c }
17:       }
18:       if (|UL| > 0) UpLevels := UpLevels + UL
19:       if (|LL| > 0) LowLevels := LL + LowLevels
20:   }
21:   return UpLevels + LowLevels  # List of Sets
22:}
```

Figure 6.  Greedy Level Decomposition algorithm

respectively, and ins($i$) as the set of the buildings that depend on $i$, and outs($i$) as the set of the buildings on which $i$ depends, and $d_{ji}$ and $d_{ij}$ are the Dedication Scores from $j$ to $i$ and the one from $i$ to $j$ respectively. For each level, all X-coordinates of the buildings are determined to minimize the sum of the energy function. Since the expression contains the Dedication Scores as coefficients, the importance of dependencies are considered in the arrangement.

*3) Block Depth Optimization:* At the third step in Fig. 4, the depth of the block (denoted as "d" in the figure) is determined. We intend that the depth of the block represents the depth of dependencies. For instance, when calls are nested deeply, the block should be deep.

To calculate the optimal depth $d$, the $d$ value that minimizes the following penalty function is searched by repeating the second step by changing $d$:

$$g(E) = \sum_{(i,j) \in E} \begin{cases} d_{ij}(|y_i - y_j| + a) & : \text{tie or reversed} \\ d_{ij}(b|y_i - y_j|) & : \text{in order} \end{cases}, \quad (2)$$

where $E$ is the set of dependencies, and $(i, j)$ denotes $i$ depends on $j$, and $y_i$ and $y_j$ are Y-coordinates of $i$ and $j$ in the block respectively. The term in "tie or reversed" case penalizes an inappropriate arrangement, i.e., the case that $i$ is lower than or equal to $j$, and the constant $a$ controls the strength of the penalty. The term in "in order" is introduced to balance the depth and the width of the block, and the constant $b$ controls the balance. In our configuration, $a$ is 2, and $b$ is 0.3.

In Fig. 4 example, the third part from the left has one tie and is penalized, but the fourth has no ties or reverses. Therefore, the fourth is better and taken.

Finally, the classes in the block are laid out. Higher levels are placed toward north, and lower levels are placed toward south. Additionally, levels are visualized as a slope as show in Fig. 7, if needed.

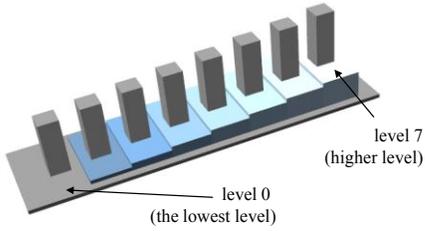

Figure 7. Slope in a City Block as a Level Representation

*E. Street Layout Algorithm for Features*

In this subsection, we explain the details of laying out features. Since the abstract tree model and the shapes of all blocks have been determined, all classes can be laid out on a 2D space in this step.

The primary goal in this step is to make relevant features placed closely. The dendrogram reflects the relevance of nodes, and the abstract tree inherits the nature. Therefore, mapping the abstract tree on a 2D space meets the goal. Besides, to minimize the distances between relevant features, the layout algorithm minimizes the following energy function:

$$h(E) = \sum_{(i,j) \in E} d_{ij}^2 \left( (x_i - x_j)^2 + (y_i - y_j)^2 \right), \quad (3)$$

where $E$ is the set of dependencies, and $(i, j)$ denotes building $i$ depends on building $j$, and $x_i$, $y_i$, $x_j$, and $y_j$ are coordinates of $i$ and $j$ respectively.

The detail of the layout algorithm is as follows:
(1) Visit the root node of the abstract tree.
(2) Place a street.
(3) For each branch child node, orthogonally lay out its subtree recursively.
(4) For each branch or leaf node, place its subtree or block along the street at the position that minimizes the energy function.
(5) Repeat from (1) while the energy function is decreasing.

After the layout, additional separator streets are placed as the blue streets shown in Fig. 2(d).

Finally, the blank SArF Map is generated in a 2D space.

*F. Information Overlaying on 2D Blank Map*

In this subsection, we explain typical representations overlaid on a SArF Map. A generated blank SArF Map is in a 2D space in order to make the third dimension freely available to various types of information. Since the blank map is used as a common mental model, it is always invariant to keep consistency in a 2D space to prevent users from confusing.

*1) Source Code Organization:* By comparing features and source code organizations such as packages, directories, and architectural documents, some important questions such as "how well is the software organized?" and "Are there any gaps between our intended architecture and the actual architecture?" can be assessed.

By mapping source code organizations to colors in SArF Map, it provides the knowledge what packages a feature consists of and how packages interact in a feature. In Fig. 8, typical color patterns found in blocks are shown. Their interpretations are as follows:

(a) *Single-color* pattern represents the feature consists of one package. If the package appears only in the feature, the package matches a single feature and is very well organized. If the package appears in several features, the feature gives the knowledge of an appropriate sub-partitioning of the package.
(b) *Layered* pattern represents the feature is implemented vertically through several packages. The pattern implies the packages are well organized. In this case, the packaging policy of the target software is layer-based, and features are not reflected in the packages and hidden. Therefore, the knowledge of the features extracted from SArF is highly valuable.
(c) *Subgroups* pattern represents the feature is implemented by several packages, or some sub-features coexist here. It may be a good chance to refactor them.
(d) *Mixed-Color* pattern represents the implementations of the feature are scattered in various packages, and it implies the packages are ill organized or the feature is cross-cutting.

Unless otherwise specified, the color of each building represents its package in all maps in this paper.

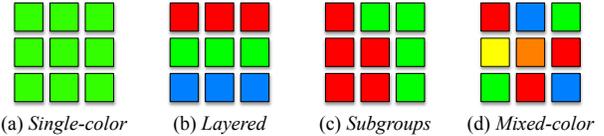

(a) *Single-color* (b) *Layered* (c) *Subgroups* (d) *Mixed-color*

Figure 8. Source Code Organization Patterns in a Block

*2) Dependency Links:* The visualization of dependency links in SArF Map is easy to understand. Since blocks are generated from a graph clustering result, most dependency links are enclosed in blocks as shown in the example in Fig. 9. Therefore, interactions between features stand out and can be easily tracked.

Dependencies or directed graph edges are displayed as curves as shown in Fig. 9. In a customary way, the curves are gradually colored from green (source ends) to red (target ends). In the figure, most intra-block curves run from north to south (i.e. green ends are upper, and red ends are lower).

To lucidly visualize inter-block dependencies, we utilized a 3D variation of the Hierarchical Edge Bundle technique [22] as previous studies [13][14].

The width of each curve represents the Dedication Score of the dependency. Therefore, important dependencies are visually emphasized.

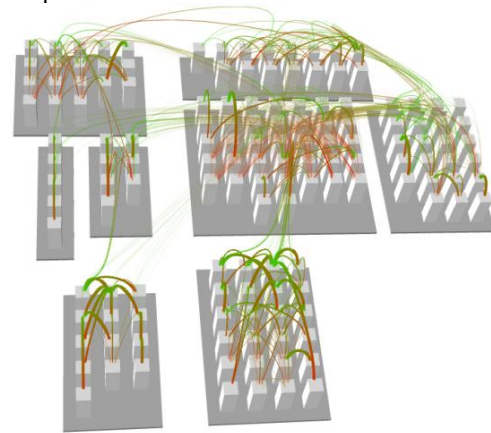

Figure 9. Link Representations in SArF Map of Weka 3.0

*3) Keywords:* Keywords characterizing each block are displayed over the block like a "tag cloud" as shown in Fig. 1. In this paper, to select keywords, the following steps are taken. (1) Words are collected from package and class names. (2) For each word, the tf-idf is calculated, where a block is regarded as a document. (3) Each word is positioned over its original class (building). (4) Words with high local density per unit area of their tf-idf values are selected to be displayed. (5) The positions of the displayed words are adjusted to avoid overlaps, if needed.

In the visualization, larger words mean that they have higher density there, i.e., more characteristic in the block.

*4) Other Properties:* Some visual attributes of the metaphors of SArF Map are freely available except the grid position of a building. Any properties and metrics needed for analyses can be mapped to them. The typical usages are summarized in TABLE II including the aforementioned usages.

TABLE II. VISUALIZED PROPERTIES MAPPING TO SARF MAP METAPHORS

| Metaphor | Attribute | Typical Used Property |
|---|---|---|
| building | color | package, category, risk |
| | height | size (LOC, #methods) |
| | shape | file types |
| | shift, rotation, tilt, texture | – |
| building ornament | brightness | utilization |
| | fire | failure, issue |
| ground | color | layer level, category |
| | height | layer level |
| link | color | link type, direction |
| | thickness | importance |

## G. Supporting System

To support SArF Map visualization, we developed two tools, a map making tool performing the procedure in Fig. 2 and a 3D viewer tool to display a generated SArF Map. The viewer tool is implemented using the OpenGL framework (http://www.opengl.org/) and provides a 3D interactive navigation, a link analysis function and a property mapping function shown in TABLE II.

## IV. RESEARCH QUESTIONS

To evaluate the effectiveness of SArF Map, we set up the following research questions. In the case studies, the questions will be assessed.

**RQ1:** Can SArF Map visualize features?
**RQ2:** Can SArF Map visualize layers?
**RQ3:** Can SArF Map be used to assess the design quality of packages?
**RQ4:** Can SArF Map reveal architectural knowledge?
**RQ5:** Can SArF Map provide consistent views for various stakeholders?

## V. CASE STUDIES

To assess the effectiveness of SArF Map, we performed five case studies from open source and proprietary software. All software systems are written in Java.

To check the correctness of the features computed by SArF Map, the correct answers from the authority of the target software are needed. We call the former the *computed features* and call the latter *the authority features*.

In these case studies, to make it easy to identify building colors, we fixed building height attributes in all SArF Maps except Fig. 16.

### A. Weka 3.0 (Open Source)

Weka is a data mining tool. The architecture of its version 3.0 is well documented in [17], and all packages other than `core` correspond to its features [23]. Therefore, they can be considered as the authority features. Figure 10(a) shows the architecture of Weka [10]. The architecture has three layers. Figure 10(b) is its SArF Map. The package color assignments of (a) and (b) are identical.

The SArF Map shows the following observations: (1) Almost blocks match the authority features in (a). It means SArF Map works effectively (RQ1 is supported). (2) In some blocks, the *Layered* patterns are found, and the observed layers match the layers in (a). It means SArF Map visualizes layers effectively (RQ2 is supported). (3) Since most features are in *Single-color* pattern or *Layered* pattern, the design of the packages is well organized (RQ3 is supported).

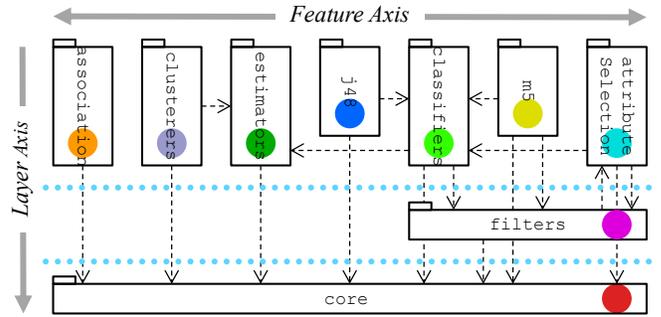

(a) Architecture of Weka 3.0 (Package Diagram)

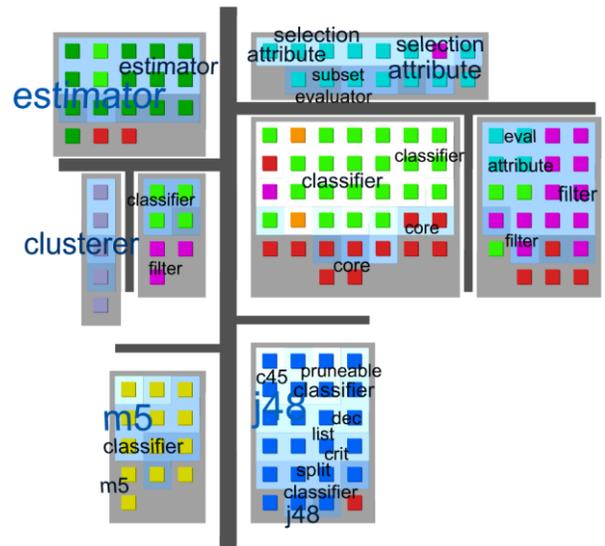

(b) Top-down view of SArF Map of Weka 3.0

Figure 10. Architecture View and SArF Map of Weka 3.0

48

## B. DMTool (Proprietary)

DMTool (tentative name) is an industrial data mining product of Fujitsu. In this case study, we had many chances to interview its developers and collected its authority features. We asked them to identify features (defined in [8]) and classes implementing the features. Its detail is in [10]. Figure 11(a) shows the architecture of DMTool. It shows the packaging policy of DMTool is layer-based. DMTool has a complicated layer structure. The right half has two layers and the left half has four layers.

Figure 11(b) and 11(c) are its SArF Maps. Keywords are not shown for reasons of confidentiality. The color assignment of (b) is identical to (a), but (c) is colored by the authority features.

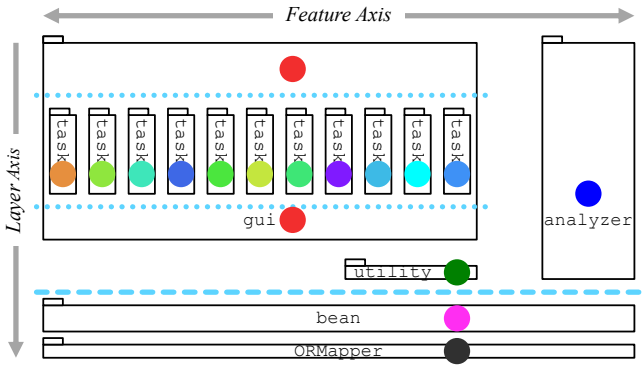

(a) Architecture of DMTool (Package Diagram)

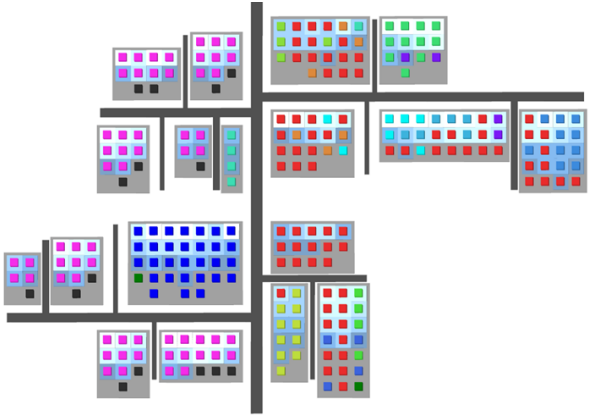

(b) Top-down view of DMTool Colored by Packages

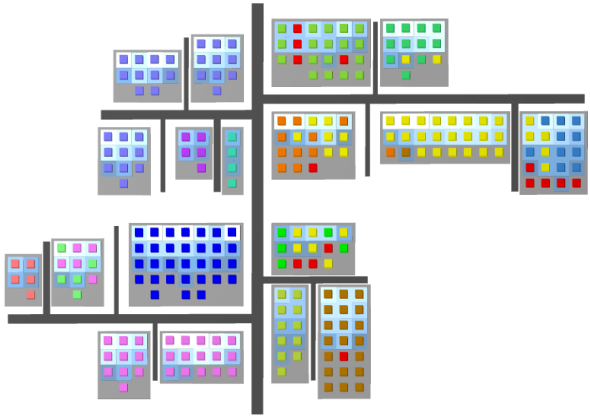

(c) Top-down view of DMTool Colored by Features
Figure 11. SArF Map of DMTool (253 classes, 16 packages, 16 features)

The observations from the two maps are as follows:

(1) In some blocks in (b), the *Layered* patterns are found, and the observed layers match the layers in (a). It means SArF Map visualizes layers effectively (RQ2 is supported). Although the layer structure of DMTool is complicated as shown in (a), SArF Map successfully visualized the layers.

(2) In (b), red classes are widely scattered in many features. It implies the `gui` package has too many features and should be divided (RQ3 is supported).

(3) In (c), most of blocks match the authority features. It means SArF Map works effectively (RQ1 is supported). Since the packaging policy is layer-based, the information of features is hidden and can be hardly obtained from the packages. SArF Map recovered such hidden features (RQ4 is supported).

(4) In the left half of (c), some neighboring blocks share the same color. It means SArF clustering algorithm excessively divided them; however, the developers said the divisions are acceptable, because the divisions match the fine-grain features of the corresponding feature they identified when they created the authority features. It also means SArF Map laid out relevant features closely.

(5) The map is divided roughly into two parts, and the division matches the architecture (RQ4 is supported).

Besides, to validate the details of the layout of the map, we interviewed the developers. Our questions were "Are the above five observations valid?", "Are the appeared source code organization patterns valid?", and "Are relevant features located close?" They answered yes to all questions.

## C. Javassist (Open Source)

Javassist is a byte code manipulation library, and Fig. 12 is the SArF Map of Javassist 3.16.1. The color assignment is auto-generated in the dictionary order of its package names so that near packages have similar colors. The same holds for the next case study.

The map shows most blocks have the *Single-Color* pattern. It implies that the packaging policy of Javassist is feature-based, and the policy is strictly kept (RQ3 is supported). The fact that the blocks highly match the packages suggests SArF Map successfully collects features (RQ1 is supported).

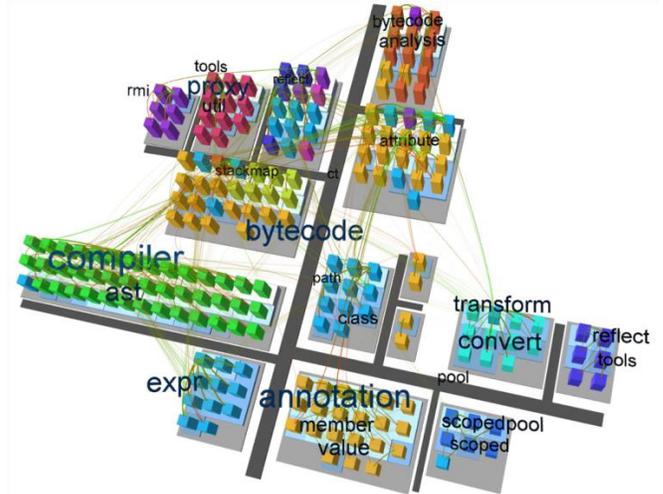

Figure 12. SArF Map of Javassist 3.16.1
(206 classes, 15 packages, fixed building height)

### D. JDK Swing (Open Source)

JDK Swing is the GUI widget toolkit of Java, and Fig. 14 is the SArF Map of JDK Swing 1.4.0. In the map the *Layered* pattern is found in most of the blocks. Some blocks have the *Single-Color* pattern. Figure 13 shows a typical dependency pattern in Swing API and its layer structure. Its layer structure (green, red, yellow and light green pattern) appears in half of the blocks. It suggests the SArF Map effectively visualizes layer structures of Swing (RQ2 is supported). Besides, it implies the packaging policy of the focused portion of Swing is layer-based.

Keywords in each block provide the knowledge what the block is. Over the blocks with the *Layered* pattern, clearly interpretable words such as "table", "border" and "combo" are observed. Therefore, it is expected that they can be named easily. We examined whether the keywords fit the features of the classes in the blocks and judged yes. It supports SArF Map effectively visualizes features and uncovers the hidden information (RQ1 and RQ4 are supported).

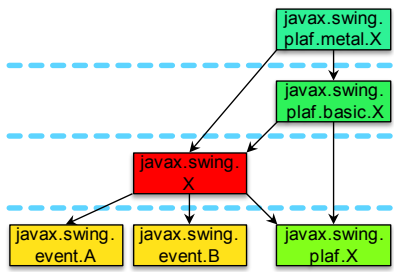

Figure 13. Typical Dependency Pattern and Layers in JDK Swing

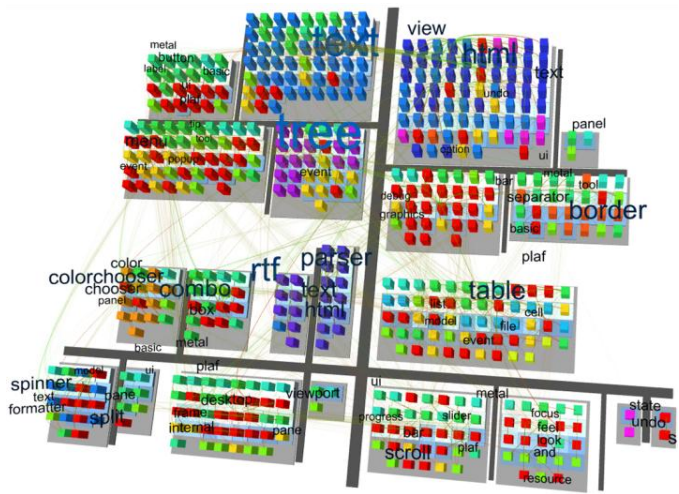

Figure 14. SArF Map of JDK Swing 1.4.0
(536 classes, 16 packages, fixed building height)

### E. An EMS (Proprietary)

In this case study, we visualize one of industrial employee management systems (EMSs) developed by Fujitsu. The aim of this case study is to assess RQ5, "Can SArF Map provide consistent views for various stakeholders?"

This case study shows two consistent maps: (1) one for non-developer decision making stakeholders and (2) another for developers in maintenance. To satisfy the needs of both stakeholders, we prepared the thoroughly different sets of properties overlaid on the same blank SArF Map.

*1) For High-level decision making:* Figure 15 is a map for high-level decision makers. In the map, maintainability, utilization and changes are represented as messiness, brightness and a building under construction respectively. Since a block means a feature, a decision maker's observation such as "a bright messy block under construction" can be interpreted as "a highly used and expanding business function with a large quality risk".

The utilization is the execution frequencies of a class or source files. It can be measured using execution traces. Besides, we took another approach in this case. We collected update operations of the database tables, and estimated the execution frequencies of classes relevant to the updates as described in [24]. The merit of this approach is it has no side effect. Therefore, it is very applicable in industrial situations.

The maintainability was estimated by using a fault-proneness prediction technique in a similar way in [25]. We predicted the fault counts of each class in the target software using its fault history and its metrics. We assigned a lower maintainability value for more predicted fault counts.

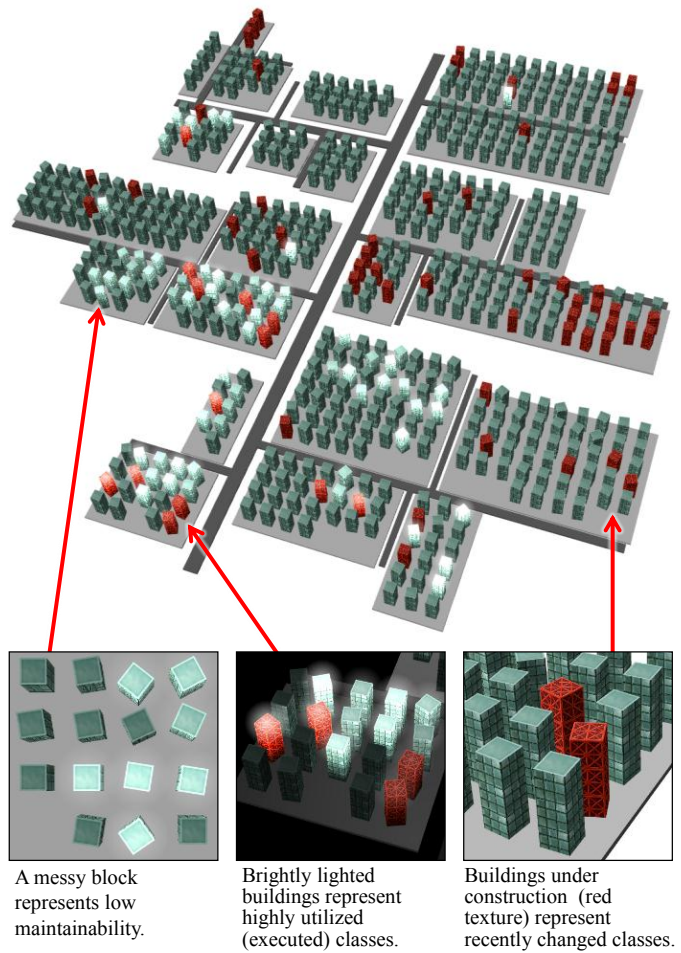

Figure 15. SArF Map of an EMS for Decision Makers
(570 classes, fixed building height)

*2) For Maintenance Tasks:* Figure 16 is the map for developers, and it shares the identical blank map with Fig. 15. It is suitable for maintenance tasks such as change impact analysis and risk assessment.

In this map, the height of a building represents the number of methods (actually, its square root is used to alleviate its right skewness), and the color represents ImpactScale [25]. ImpactScale is a metric quantifying the scale of change impact and is used in risk, effort and quality assessment. High ImpactScale values mean risky and are visualized in red color. Low ImpactScale values are visualized in unnoticeable gray color. Dependency links are also displayed.

In Fig. 16, we focus on the yellow class indicated by an arrow, and then all dependency links from and to the class are high-lighted by our 3D viewer tool. In such a way, user can assess the change impact easily.

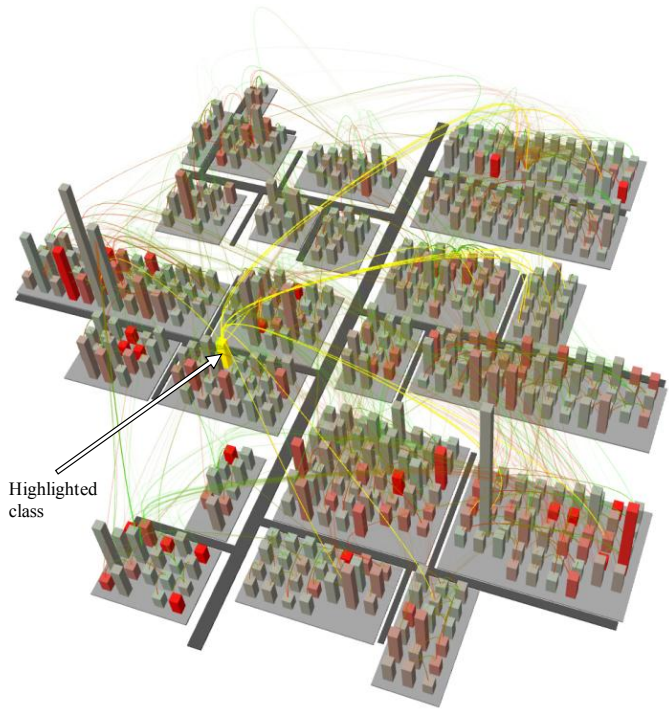

Figure 16. SArF Map of an EMS for Software Maintenance Engineers (height: number of methods, color: ImpactScale [25])

*3) Results:* Although both the two maps provide different information, they share the identical blank map. Therefore, decision makers and developers can communicate with the same mental model.

One of the use cases for maintenance developers is that they negotiate to acquire the budget for preventive maintenance by sharing an awareness of risky portions of the system. Presently, the effectiveness of such use cases using SArF Maps is still under review. As a result, RQ5 is supported, but how it facilitates communication between different stakeholders are to be assessed in future work.

To share the mental model, we found naming features was a large issue for the decision makers' understanding. The extracted keywords by the current naïve keyword selection technique were not satisfiable for them, and the developer had to refine the keywords into understandable feature names.

*F. Summary of Research Questions*

As already stated, all research questions are supported in the case studies. To generalize the results and to evaluate how effective the visualization is in practical situations, further experiments and analyses should be investigated.

## VI. THREATS TO VALIDITY

There are some threats to the validity of our proposal and results.

The most important threat is the hypothesis that features are successfully collected by SArF clustering algorithm. The original study presented some evidences [10], and we visually confirmed them again in this paper. However, further case studies and experiments are needed to validate the hypothesis. Even if features are successfully collected by the algorithm, it is necessary to note that features can overlap and various features from various viewpoints can coexist. The current algorithm does not support such situations.

Another threat is the quality of dependency graph. The quality of the map depends on the quality of the extracted dependency graph. If reflections and dependency injections are heavily used in target software, an incomplete map is generated.

To visualize larger software systems, two scalability problems arise. The first problem is granularity. When a system with over several thousands of classes is visualized, too large city blocks (with over several hundreds of classes) appear and are difficult to be managed.

The second scalability problem is color resolution. Since human cannot discriminate too many colors, when packages are too many, source code organization patterns are difficult to identify.

## VII. SUMMARY AND FUTURE WORK

*A. Summary*

We have proposed a novel software visualization technique, *SArF Map*. It visualizes software architecture from a feature and layer viewpoints using a city metaphor. SArF Map visualizes implicit software features extracted by our previous study, SArF software clustering algorithm that gathers features.

Since features are high-level abstraction units in software systems, it is expected that users of SArF Map can make high-level decisions on the target system. Besides, it is interpretable and can be used as a common mental model of the target system among various statuses of stakeholders.

To visualize two viewpoints, features and layers, SArF Map uses the *Street Block Tree Layout*. In the layout, classes are represented as buildings, and features are as city blocks, and layers are as slopes and north-south direction. Relevant features are connected by streets and located in close positions in order to facilitate the understanding of the architecture from a feature viewpoint. The layout algorithms find the optimal layout using the energy minimization strategy.

There are two main contributions in this paper. The first is combining dependency-based software clustering and software visualization. Since the clustering algorithm extract features and untangle complicated layer structures, the second contribution, decomposing and visualizing layers are achieved. The combination of features and layers has synergistic effect. For example, by comparing package colors with features and layers in a SArF Map, the source code organization patterns of packages can be identified.

A blank SArF Map is generated in a 2D space in order to make the third dimension freely available to various types of information. Keywords extracted from features are displayed like tag clouds. Since SArF Map uses a graph clustering, only inter-feature dependency links visually stand out and are easy to understand.

SArF Map is fully automated, and the output is deterministic, and the input is only jar files or a dependency graph. It is language-independent. These characteristics enable SArF Map highly applicable in various industrial situations.

Case studies using OSS and industrial software show that SArF Map successfully visualizes features and layers and can uncover architectural knowledge and that it is also used for various stakeholders.

*B. Future Work*

For future work, further detailed investigation of findings of SArF Map and the qualitative and quantitative evaluations of the effectiveness should be performed, because the case studies are preliminary evaluations and are mainly focused on confirming whether SArF Map properly work.

The aforementioned issues such as scalability, keyword selection and the quality of dependency graph should be further explored. To extract dependency graph with less false positives and negatives, combining a semantic analysis and a dynamic analysis with a static code analysis is promising.

To visualize software evolution like previous studies [7][11], a map layout algorithm should be robust to minor changes of software. Although SArF clustering algorithm is enough robust to changes [10], the current layout algorithm is not very robust. We plan to develop a more robust layout algorithm.

REFERENCES

[1] S. Ducasse and D. Pollet, "Software architecture reconstruction: a process-oriented taxonomy," *IEEE Trans. on Softw. Eng.*, vol. 35, no. 4, pp. 573-591, 2009.

[2] A. R. Teyseyre and M. R. Campo, "An overview of 3D software visualization," *IEEE Trans. on Vis. Comput. Graph.*, vol. 15, no. 1, pp. 87–105, 2009.

[3] P. Caserta and O. Zendra, "Visualization of the static aspects of software: A survey," *IEEE Trans. on Vis. Comput. Graph.*, vol. 17, no. 7, pp. 913-933, 2011.

[4] G. Langelier, H. Sahraoui, and P. Poulin, "Visualization-based analysis of quality for large-scale software systems," *Int'l Conf. on Automated Softw. Eng.*, ASE, p. 214-223, 2005.

[5] R. Wettel and M. Lanza, "Visualizing software systems as cities," *Int'l Workshop on Visualizing Software for Understanding and Analysis*, VISSOFT, pp. 92–99, 2007.

[6] T. Panas, T. Epperly, D. Quinlan, A. Saebjornsen, and R. Vuduc, "Communicating software architecture using a unified single-view visualization," *Int'l Conf. on Engineering Complex Computer Systems*, ICECCS, pp. 217-228, 2007.

[7] F. Steinbrückner and C. Lewerentz, "Representing development history in software cities," *ACM Sympo. on Software Visualization*, SoftVis, pp. 193-202., 2010.

[8] T. Eisenbarth, R. Koschke, and D. Simon, "Locating features in source code," *IEEE Trans. on Softw. Eng.*, vol. 29, no. 3, pp. 210-224, 2003.

[9] C. Knight and M. Munro, "Virtual but visible software," *Int'l Conf. on Information Visualization*, InfoVis, pp. 198-205, 2000.

[10] K. Kobayashi, M. Kamimura, K. Kato, K. Yano, and A. Matsuo, "Feature-gathering dependency-based software clustering using dedication and modularity," *Int'l Conf. on Softw. Maint.*, ICSM, pp. 462-471, 2012.

[11] A. Kuhn, P. Loretan, and O. Nierstrasz, "Consistent layout for thematic software maps," *Working Conf. on Rev. Eng.*, WCRE, pp. 209–218, 2008.

[12] J. I. Maletic, A. Marcus, G. Dunlap, and J. Leigh, "Visualizing object-oriented software in virtual reality," *Int'l Workshop on Prog. Compre.*, IWPC, pp. 26–35, 2001.

[13] P. Caserta, O. Zendra, and D. Bodenes, "3D hierarchical edge bundles to visualize relations in a software city metaphor," *Int'l Workshop on Visualizing Software for Understanding and Analysis*, VISSOFT, pp. 1-8, 2011.

[14] R. Wettel, "Software systems as cities," PhD Thesis, University of Lugano, 2010.

[15] S. Mancoridis, B. S. Mitchell, Y. Chen, and E. R. Gansner, "Bunch: a clustering tool for the recovery and maintenance of software system structures," *Int'l Conf. on Softw. Maint.*, ICSM, pp. 50-59, 1999.

[16] G. Scanniello, A. D'Amico, C. D'Amico, and T. D'Amico, "Architectural layer recovery for software system understanding and evolution," *Software: Practice and Experience*, vol. 40, no. 10, pp. 897-916, 2010.

[17] H. Witten and E. Frank, "Data Mining Practical machine learning tools and techniques," Morgan Kaufmann, 2005.

[18] M. Newman, "Fast algorithm for detecting community structure in networks," *Physical Review E*, vol. 69, no. 6, pp. 1-5, 2004.

[19] A. Clauset, M. Newman, and C. Moore, "Finding community structure in very large networks," *Physical Review E*, vol. 70, no. 6, 2004.

[20] E. A. Leicht and M. E. J. Newman, "Community structure in directed networks," *Physical Review Letters*, vol. 100, no. 11, p. 118703, 2008.

[21] I. G. Tollis, G. Di Battista, P. Eades, and R. Tamassia, "Graph drawing: Algorithms for the visualization of graphs," Prentice Hall, 1998.

[22] D. Holten, "Hierarchical edge bundles: Visualization of adjacency relations in hierarchical data," *IEEE Trans. on Vis. Comput. Graph.*, vol. 12, no. 5, pp. 741-748, 2006.

[23] C. Patel, A. Hamou-Lhadj, and J. Rilling, "Software clustering using dynamic analysis and static dependencies," *Euro. Conf. on Softw. Maint. and Reeng.*, CSMR, pp. 27-36, 2009.

[24] K. Kato, T. Kanai, and S. Uehara, "Source code partitioning using process mining," *Int'l Conf. on Business Process Management*, BPM, LNCS 6896, pp. 38-49, 2011.

[25] K. Kobayashi, A. Matsuo, K. Inoue, Y. Hayase, M. Kamimura, and T. Yoshino, "ImpactScale: Quantifying change impact to predict faults in large software systems," *Int'l Conf. on Softw. Maint.*, ICSM, pp. 43-52, 2011.